# Three-step Formation of Diamonds in Shock-compressed Hydrocarbons: Decomposition, Species Separation, and Nucleation

Bo Chen[1], Qiyu Zeng[1], Xiaoxiang Yu[1], Jiahao Chen[1], Shen Zhang[1], Dongdong Kang[1], Jiayu Dai[1, *]

[1]Department of Physics, National University of Defense Technology, Changsha 410073, P. R. China

*Corresponding authors. E-mail addresses: jydai@nudt.edu.cn

## ABSTRACT

The accumulation and circulation of carbon–hydrogen dictate the chemical evolution of ice giant planets. Species separation and diamond precipitation have been reported in carbon–hydrogen systems, verified by static and shock-compression experiments. Nevertheless, the dynamic formation processes for the above-mentioned phenomena are still insufficiently understood. Here, combining deep learning model, we demonstrate that diamonds form through a three-step process involving decomposition, species separation and nucleation procedures. Under shock condition of 125 GPa and 4590 K, hydrocarbons are decomposed to give hydrogen and low-molecular-weight alkanes ($CH_4$ and $C_2H_6$), which escape from the carbon chains resulting in C/H species separation. The remaining carbon atoms without C–H bonds accumulate and nucleate to form diamond crystals. The process of diamond growth is found to associated with a critical nucleus size where dynamic energy barrier plays a key role. These dynamic processes for diamonds formation are insightful in establishing the model for ice giant planet evolution.

Hydrocarbons are one of the most abundant molecules in the universe, together with water and ammonia, consisting the dominant components of the mantles over ice giant planets such as Neptune and Uranus[1-4]. A plethora of chemical mixtures can be generated under extreme conditions of hundreds of GPa and thousands of K in ice giant planets, which implies a rather complicated chemical and physical processes occurring in deep interiors[5-7]. In contrast with terrestrial and gas planets, the complex dynamical process within the ice giant planet features in the thermal evolution[6, 8], magnetic field distribution[9, 10], and electrical properties[2]. Under extreme conditions, the state of matter will experience into an intermediate state between condensed matter and plasma[11], which leads to the identification of novel features[12, 13] in normal materials[14], such as, mixtures of ammonia and water undergo auto-ionization and are found to be superionic at high pressures[15-18], Methane and saturated hydrocarbons undergo decomposition and polymerization reactions forming diamonds or metallic hydrogen[5, 7, 19-21]. The existence of conductive compounds is likely resulting in the development of a planetary dynamo, which could be responsible for the unusuality of magnetic fields in Uranus and Neptune[5, 9, 10]. Thus, the investigation of the chemical dynamics for hydrocarbons under extreme conditions is vital in further understanding the ice giant planets in the universe.

Quite a few of experimental trials have been made to investigate the decomposition of hydrocarbons in the laboratory. Considerable differences are found in the dissociation pressure thresholds realised in static and shock compression experiments. No decomposition was found up to 288 GPa at room temperature in diamond anvil cells[22, 23]. On the contrary, some evidence for polymerization was found at around 1100 K by laser heating and additional proof of dissociation and diamond formation was reported at 3000 K under 10–80 GPa[24, 25]. These clues highlight the heating as the dominant factor that affects the chemical and physical processes occurring in static compression experiments[26]. Phase separation into diamond and hydrogen was found in laser-driven shock compression experiments with polystyrene $(C_8H_8)_n$ samples using the advanced in-situ X-ray diffraction method, which, meanwhile, requires pressures up to 150 GPa (about 10 times higher than the pressure used in static compression

experiments[20]). Hydrocarbon decomposition is tightly correlated with the formation of chemical bonds and the chemical reaction environment[27]. The results of diamond anvil cell experiments may be influenced by other factors that affect the chemical processes involved, such as absorption of molecular hydrogen by the diamond anvils, which promotes separation[20]. In a word, by reviewing the above-mentioned results, the understanding of the dynamic processes of hydrocarbons under extreme conditions is still insufficient that calls for further in-depth investigation through theoretical considerations.

Numerous first-principles calculations have been performed to explore C–H phase diagrams and potential stable compounds[5, 28]. According to *ab initio* molecular dynamics calculations, pressures >300 GPa and a temperature up to 4000 K are required for methane to decompose eventually to produce diamond[19], during which the liquid structures have been observed in hydrocarbons under the thermodynamic conditions used in shock experiments[29]. Some clues of C-H species separation can be found at extreme conditions such as in the inertial confinement fusion,[30] but the direct observation of species separation and diamond formation has not been realised in *ab initio* molecular dynamics studies of shock-compressed hydrocarbons due to the limitations of simulation time and size[29]. Furthermore, this dynamic process analogous to the incidence occurred in planet interiors is poorly understood from the perspective of calculation, albeit the theoretical studies are valuable and insightful for predicting the static thermodynamic properties and phase boundaries of hydrocarbons. The atomistic mechanisms involved in species separation and diamond formation after shock compression remain elusive so far. Here, we performed large-scale molecular dynamics (MD) simulations based on the reactive force field (ReaxFF) and machine-learning-based potential and investigated the processes of polyethylene (PE) during shock compression. At ~125 GPa and 4590 K, PE was found to decompose into smaller molecules ($H_2$, $CH_4$, and $C_2H_6$), and the diamond nucleation process was explicitly observed in carbon-rich regions after the separation. We also report the spontaneous growth of a diamond crystal requiring a critical nucleation radius, which agrees with the classical nucleation theory. Our findings reform the existed perspectives of the chemical and physical processes of the phase separation involving the evolution from hydrocarbons to diamond and hydrogen, which could assist the establishment of the evolutionary model for ice giant planets.

## Results

**Shock-compressed hydrocarbons.** Rapid compression drives the materials into extreme thermodynamic conditions that related to the initial states according to Ranke-Hugoniot relationship. We first validated the accuracy of the ReaxFF potential[31, 32] in describing the shocked states of PE by using the multi-scale shock technique (MSST)[33] (see details in the methods section). The Hugoniot curves in the $u_s$–$u_p$ and $P$–$V/V_0$ planes calculated by MD simulations fit the experimental results quite well[34] (Fig. S1). Considering the shocked structures obtained from the single-shock-compression simulations, a three-step transformation path of pristine polymers is identified, decomposition, species separation, and diamonds nucleation (Fig. 1). The crystalline PE firstly melt under shock loading while the hydrocarbon chains are not broken with the shock velocity $U_s$ below 12 km/s, where the temperature $T$ is ~2330 K and the pressure $P$ is ~72 GPa. By increasing the shock velocity to reach higher $T$ and $P$ along the Hugoniot curves, the decomposition of hydrocarbons into small molecules is observed without C/H separation until reaching the phase separation boundary. This regime includes the single-shock with velocity $U_s$ corresponding to 13 km/s ($T$~2919 K, $P$~88 GPa) and 14 km/s ($T$~3534 K, $P$~106 GPa), and double-shock compression from the first stage of 11 km/s to the second stage of 16 km/s ($T$~2715 K, $P$~150 GPa). When the shock velocity increases up to 15 km/s ($T$~4590 K, $P$~125 GPa), the hydrocarbons dissociate dramatically, and followed by the species separation into a hydrogen-rich and a carbon-rich region. Such species separation is also found in the shocked amorphous PE in the same shock states (Fig. S2). The process of separation lasts for about 200 ps, and a nucleus of diamond forms within the carbon-rich region, gradually growing to a sizeable diamond crystal. This three-step formation process of diamonds has also been observed at higher pressures by the double-shock-compression up to 254 GPa. These MD results are convincingly in accords with the trend reported by previous experiments[20] and *ab initio* calculations[28]. Moreover, from atomic scale perspective, we propose a three-step transition pathway from hydrocarbons to diamonds following the Hugoniont curves, which provides a profound understanding for the evolutionary process under the conditions in deep interiors of ice giants.

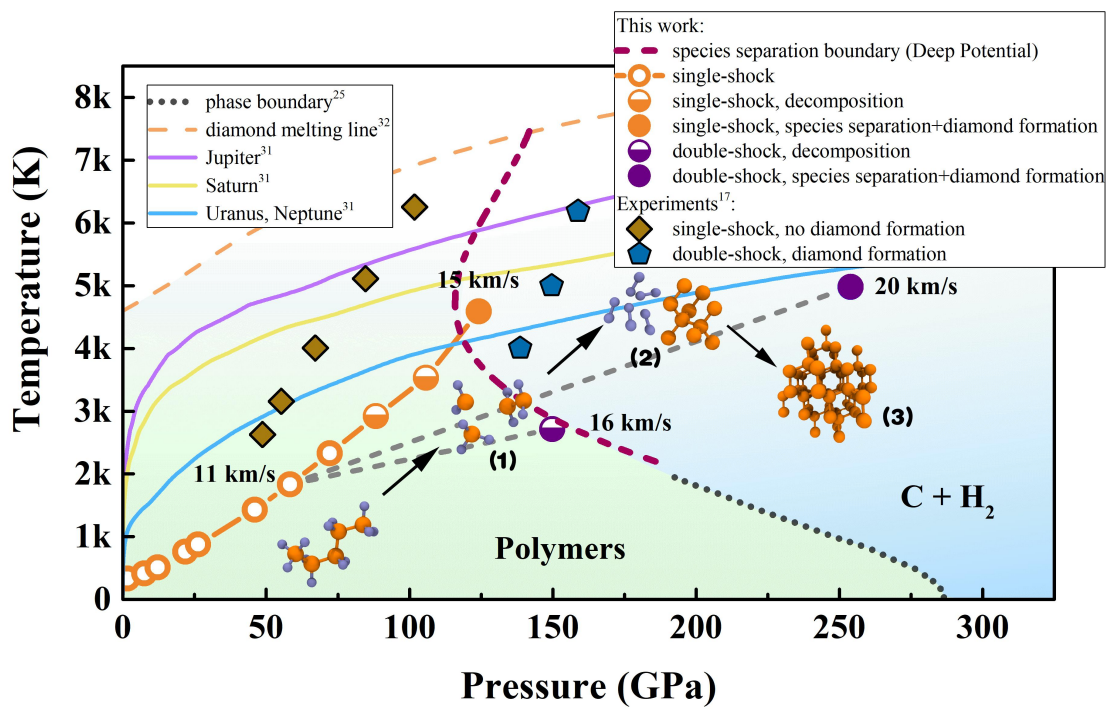

**Fig. 1. Summary of thermodynamic conditions and formation processes of diamonds in Shock-compressed Hydrocarbons.** The inserted schematic diagrams of atomic structures (orange particles for carbon atoms and iceblue particles for hydrogen atoms) represent the three-step formation of diamonds: (1) decomposition of hydrocarbon polymers into molecules, (2) C/H species separation, and (3) nucleation of diamonds. These agree with the results given by the laser-induced shock experiments[20]. The predicted isentropes of Jupiter[35], Saturn[35], as well as Uranus and Neptune[35] cross the polymers/C+$H_2$ phase boundary[28] and locate below the diamond melting line[36].

**Decomposition.** As shown in Fig. 2(a), after shock compression, the radial distribution function (RDF) curves are broadened extraordinarily, indicating that the initial crystalline solid melting into a liquid phase. The liquid structures of shock-compressed hydrocarbons have also been observed in experiments and by performing theoretical calculations[29, 37, 38]. When the shock velocity approaches beyond 12 km/s, the infused shock energy exceeds the threshold of bond energy in hydrocarbons, and the decomposition starts in $-[-CH_2-]_n-$ chains, generating small molecules. The most obvious feature in this process is the production of $H_2$ molecules, where a new characteristic peak at 0.72 Å is found in the RDF curve of $g_{H-H}(r)$. In addition, the height of the characteristic peak augments with the increase of shock velocity, indicating the occurrence of a more intense decomposition-reaction. Besides hydrogen, a series of small molecules are also found in the product, and the compositions of these molecules at different shock velocities are obtained by virtue of bond orders analysis. As presented in Fig. 2(b), among all the products, $H_2$, $CH_4$ and $C_2H_6$ molecules serve as the dominant components. These molecules exhibit lower C/H ratios than that of pristine PE (1:2), and the decomposition to those hydrogen-rich molecules leads to a significant decrease of hydrogen content in the remaining $C_xH_y$ chains. Due to the intense decomposition-reaction caused by the increase in shock strength, more hydrogen-rich molecules are generated, further reducing the hydrogen content in the remaining chains (inset of Fig. 2b). In particular, at the shock velocity of 15 km/s, the quantities of carbon atoms in the remaining chains exceed that of hydrogen atoms, where the species separation occurs. These results demonstrate that the decomposition reactions play an important role as the precursor of C/H species separation.

To figure out the details of decomposition process at the shock velocity of 15 km/s, we investigate the temporal evolution of produced molecules as presented in Fig. 2(c). The number of produced molecules surges up in the initial 50 ps, reaching to about 72% of the deposition-reaction yields, and then the productivity slows down in the following hundreds of picoseconds. As hydrogen-rich molecules release due to decomposition, the C/H ratio of the original molecular chains increases with a maximum of C:H=1.8 at 450 ps. The temporal evolution of the thermodynamic quantities is accordingly investigated, as shown in supplementary Fig. S4. Pressure $P$, density $\rho$, and shock energy $E_{msst}$ reach equilibrium after 50 ps, indicating that the entire system has arrived to a steady shock state, while temperature $T$ continues increasing slowly even without shock energy being applied (i.e., at a stable $E_{msst}$). This can be explained by the formation of chemical bonds, which releases heat and further warms up the temperature of the system. Thus, the decomposition reactions are vigorous within 50 ps of the shock loading, and the chemical processes is persisting, however, gentle in the following few hundred picoseconds.

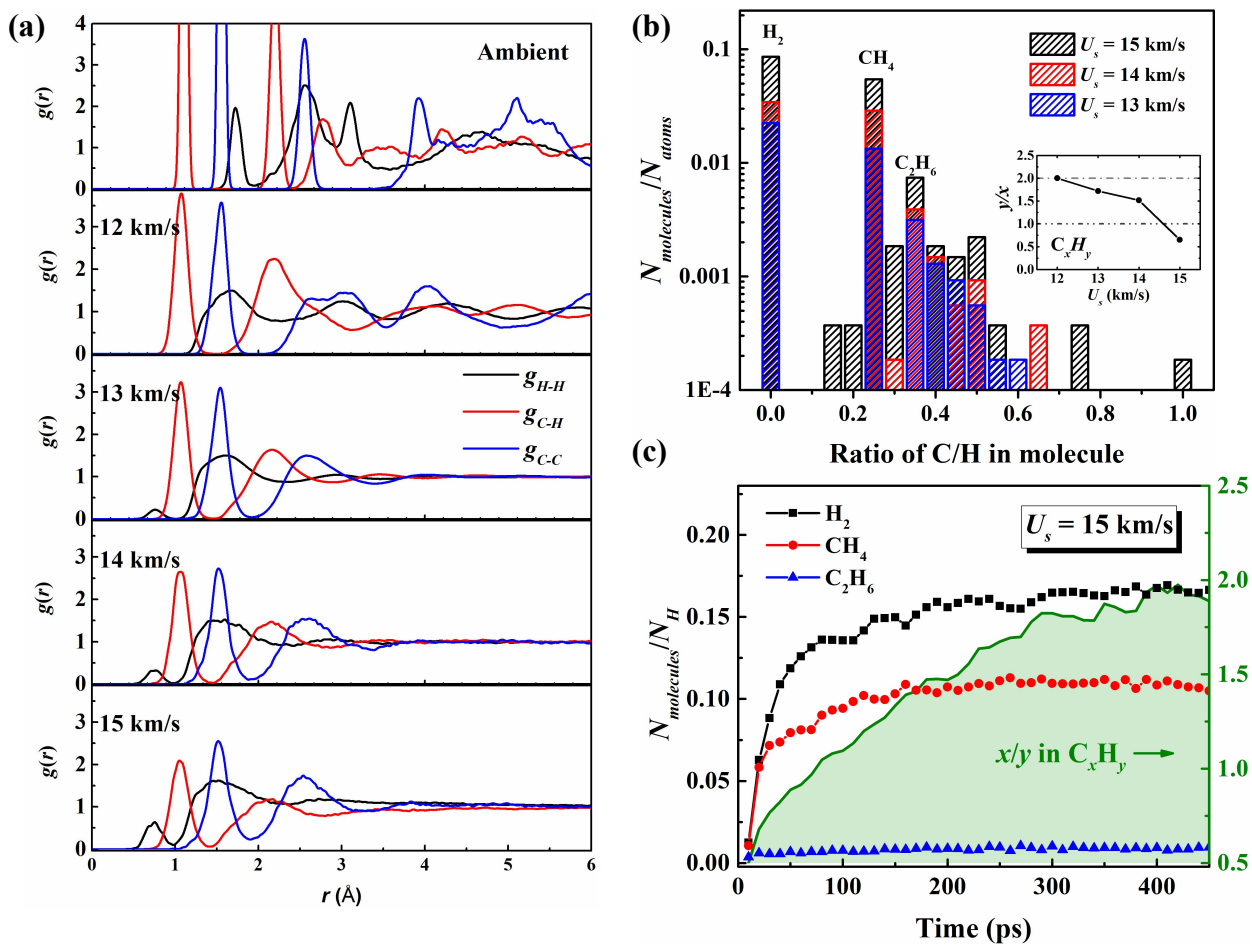

**Fig. 2. Molecules produced in shock-compressed PE.** (a) Radial distribution functions of polyethylene at ambient conditions and under shock compression. (b) Compositions of molecular products after reaching reaction equilibrium at shock velocities of 13, 14, and 15 km/s in the system containing 1800 carbon atoms and 3600 hydrogen atoms. The inset indicates the H/C ratio of the remaining chains. (c) Temporal evolution of the ratio of produced molecules to the total number of hydrogen atoms in PE subjected to shock waves at a velocity of 15 km/s.

**Species separation.** At the shock velocity of 15 km/s, with a temperature of ~4590 K and a pressure of ~125 GPa yield, the thermodynamic condition reaches a C/H separation regime. As shown in Fig. 3, at 50 ps after shock loading, C and H atoms are homogeneously distributed in the space. To quantitatively estimate the extent of separation, we sum the number of hydrogen atoms $n(H)$ and carbons atoms $n(C)$ in the *x-z* plane, and calculate the species ratio with $n(H)/n(C)$ for each area in the space. The distribution of $n(H)/n(C)$ centralizes around 2, corresponding to the value of pristine PE, which proves the homogeneously mixing state. Meanwhile, massive C-H bonds rupture occurs in the system, while the number of H-H bonds surges up, forming a clear characteristic peak of hydrogen in the RDF of $g_{H-H}$. During this period, the decomposition dominates the process, and there is little aggregation of the generated hydrogen molecules. In the subsequent few hundred picoseconds, the atoms gradually separate into carbon-rich region and hydrogen-rich region, and eventually form a well-defined boundary between separated C and H atoms at 225 ps. Besides, the distribution of $n(H)/n(C)$ is significantly broadened and dominantly weighted at the values near 0 and greater than 5, illustrating the system has reached a large extent of separation to C and H. The formation rate of H-H bonds is much lower in this circumstance, therefore most of the hydrogen molecules are produced due to the decomposition reactions within 50 ps, mixing homogenously with other compositions. After decomposition, the produced hydrogen-rich molecules leave from the original PE chains due to the movement, and the carbon atoms that lose coordinated hydrogen atoms bond to each other, promoting further separation in the system.

Considering these two processes, the separation of species is dominated by two groups of factors. On one hand, the thermodynamic conditions to activate the decomposition reactions (when C–H chains split into small molecules) serve as the prerequisite of separation. The dynamics process that hydrogen-rich molecules, especially for hydrogen molecules, move away from the remaining CH chains, stimulate the C-C bonding for further separation. Thus, strong shock will cause a violent decomposition to generate plenty of small molecules, which can spontaneously escape from the chains, finally achieving the separation. On the other hand, if the hydrogen-rich molecules can be manipulated to leave the mixtures of hydrocarbons, species separation can also be achieved even under thermodynamic conditions with low yields. This may explain why separation occurs more readily in diamond anvil cell experiments than in other experiments, since the presence of diamond or other substrates introduces an externally forced process such as molecular absorption, which promotes movement of hydrogen molecules and requires less extreme thermodynamic conditions. This is a reasonable explanation for the discrepancy between the pressure thresholds for hydrocarbon decomposition in static and shock compression experiments.

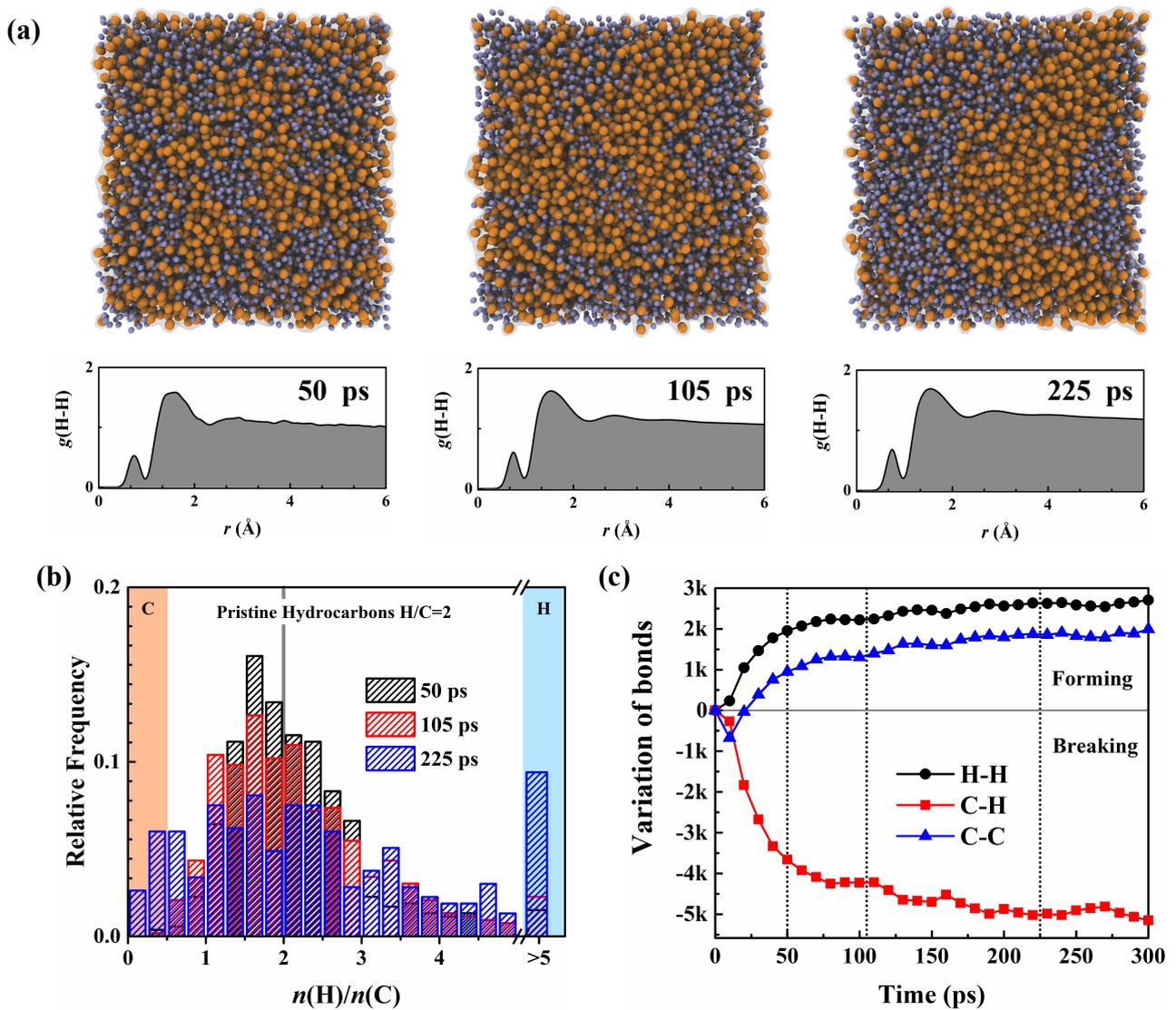

**Fig. 3. Processes of C/H species separation under conditions with shock velocity of 15 km/s.** (a) The snapshots shock-compressed PE at different simulation times. Below are their corresponding pair correlation function of H-H. Orange spheres denote carbon atoms and iceblue spheres denote hydrogen atoms. (b) Relative frequency of the number ratio of hydrogen atoms and carbon atoms. The number of $n$(H) and $n$(C) was obtained in the *x-z* spatial plane with bins of 2 Å×2 Å. (c) The temporal evolution of bonds variation in the system. Dot lines indicate the bonds information of the corresponding three moments.

**Dynamical behaviors of hydrogen.** To further verify the results of species separation and investigate the behaviors of hydrogen in H-rich regions, we develop a deep-neural-network-based potential (DP) to construct the PES with the accuracy of *ab initio* calculations (details are listed in the methods section). The performance of the DP is shown in supplementary Table 1 and supplementary Fig. S5. It is clear that the DFT energies can be accurately reproduced by the DP among different components of hydrocarbon systems, including pure C, pure H, CH, $CH_2$, and other proportions of C/H. The mean absolute errors are no more than 0.05 eV/atom. The DP is primarily validated by the comparison of RDF in the systems of pure hydrogen, pure carbon and $CH_2$, and the Hugonoit loci in the *P-T* plane. These MD results are in agreement with DFT calculations and the trend reported by shock experiments.

A series of MSST simulations of PE with shock velocities from 4-18 km/s are performed by using DP, and the species separation is firstly observed at the same shock velocity 15 km/s. Fig 4(a) shows the comparison of RDF of shock-compressed PE between DP and ReaxFF. The pair correlation function of C-C and C-H are almost consistent, but the $g_{H-H}$ of DP shows a continuous and wide peak, enveloping two independent peaks of the molecular and atomic hydrogen in ReaxFF. The results of DP indicate that the hydrogen exists as a mixture state of molecular and atomic phase after species separation. As presented in Fig 4(b), $H_2$ peak appears and grows at first 20ps, along with the hydrogen aggregation locally. Subsequently, abundant bonds form among carbon atoms, and a noticeable aggregation of carbon atoms begins to appear after 50ps because of the loss of hydrogen atom coordination. Over 300ps after shock compression, we explore the dynamical behaviors of bonded hydrogen atoms in H-rich region. When the distance between hydrogen atoms is less than 0.72 Å, these two atoms is considered to form a bond. As shown in Fig. 4(c), the hydrogen bonds do not exit stably, and the formation and break of hydrogen bonds dynamically and frequently in the system. It is the reason why $g_{H-H}$ of DP shows as a characteristic peak of enveloping molecular and atomic hydrogen. With exploring different thermodynamic conditions (details are presented in supplementary Fig. S6), we use DP to obtain a phase boundary of species separation as shown in Fig. 1, in agreement with the results of experiments and calculations from ReaxFF. Verified by the method with *ab initio* accuracy, it can be concluded that the evolution path given by ReaxFF is valid, but the stability of the H-H bond is overestimated, so that the dynamical transition between atomic and molecular hydrogen is missing.

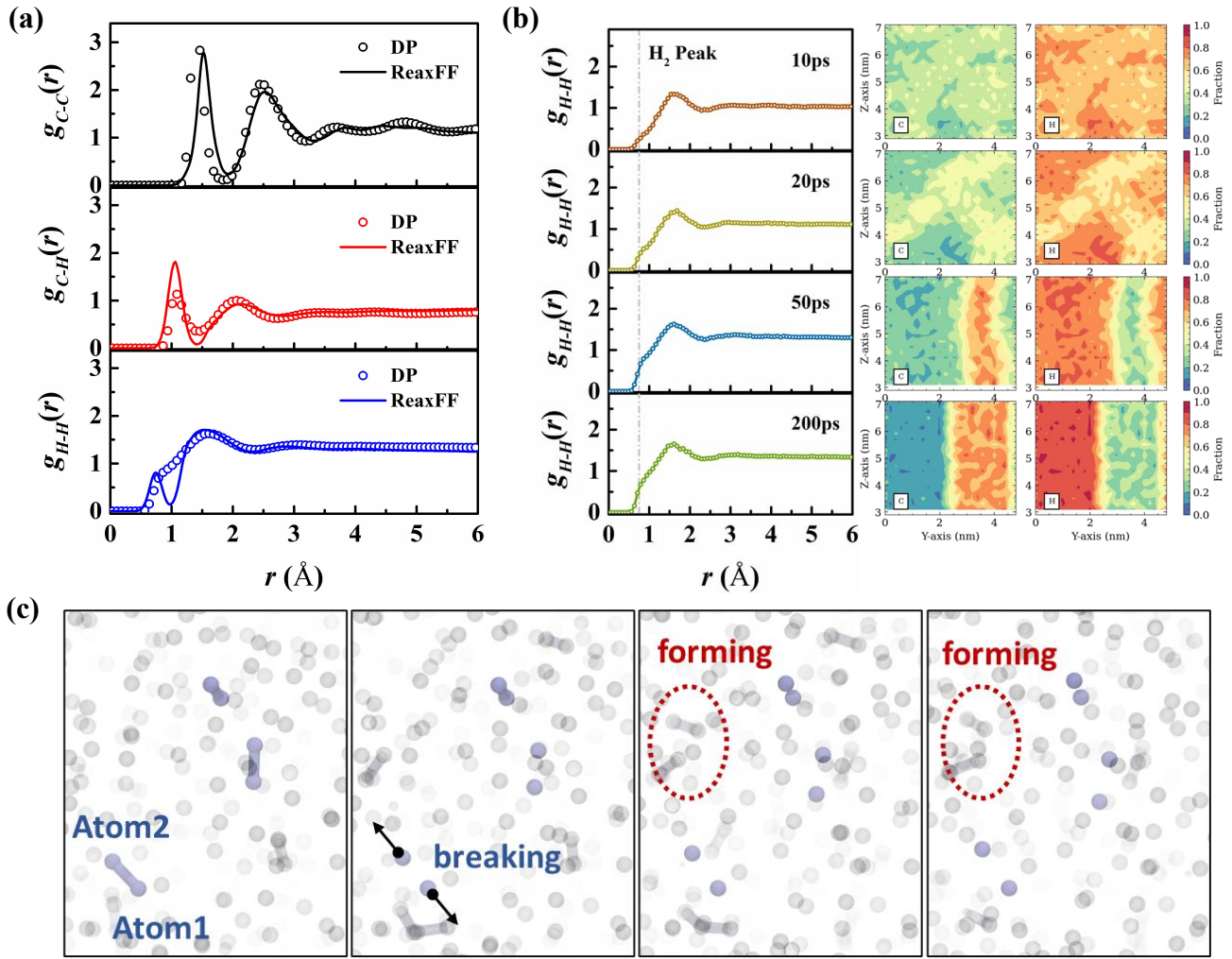

**Fig. 4. Shock simulations by using Deep potential.** (a) Temporal evolution of pair correlation function of H-H and the spatial distribution of elements at the corresponding time. (b) Radial distribution functions of shock-compressed polyethylene by using ReaxFF and DP. (c) The snapshots of H-H bonds break and formation in the H-rich region.

**Nucleation of diamonds.** As the separation proceeds, carbon atoms gain more carbon-coordination and accumulate into one region. To identify the structure types of these atoms, we perform a detailed structural analysis using the extended common neighbour analysis method (as described in the methods section). With the aid of diagnosis method, carbon atoms that form a diamond-crystal structure in the disordered system can be identified. As shown in Fig. 5(a), the dynamic diamond formation and nucleation are observed in the system containing 28800 atoms. At 235 ps, a diamond-crystal embryo forms for the first time, and crystalline embryos grow stochastically but vanish quickly in the subsequent tens of picoseconds. A stable nucleus forms at 270 ps and eventually starts to grow into a large crystal at an extremely high growth rate of around 63.318 μm/h (Fig. 5b). The diamond nucleation process is presented in a video available in the supplementary material. In addition, we also calculate the X-ray diffraction spectra of shock-compressed PE at different shock times to compare with the results of experiments. At the beginning of shock, there is a clear signal for amorphous PE and a broad peak centered on 2.2 Å$^{-1}$, indicating the presence of compressed C–H liquid. A new diffraction feature at 3.02 Å$^{-1}$, compatible with the Bragg reflections of compressed diamond, emerges at 450 ps, which agrees well with the results of recent shock experiments. These are the first theoretical results of the observation of dynamic processes in diamond formation from shock-compressed hydrocarbons. To investigate the effect of computational size on diamond nucleation, a small system containing 5400 atoms is replenished to simulate shock compression under the same condition (Fig. S7). The temporal evolution of volume in diamond structure in Fig. 5(b) indicates that only unstable crystalline embryos form in the 5400-atom system and there is no sign of nucleation. The classical nucleation theory[39, 40] is applied to describe the diamond nucleation behaviour (as described in the methods section). A minimum of $N_c$ atoms is required in the solid nucleus to achieve liquid–solid phase transition. At this shock condition, the critical nucleus size $N_c$ is estimated as ~115. This indicates the reason why diamond formation cannot be observed in small systems by performing density functional theory calculations until $N_c$ is met.

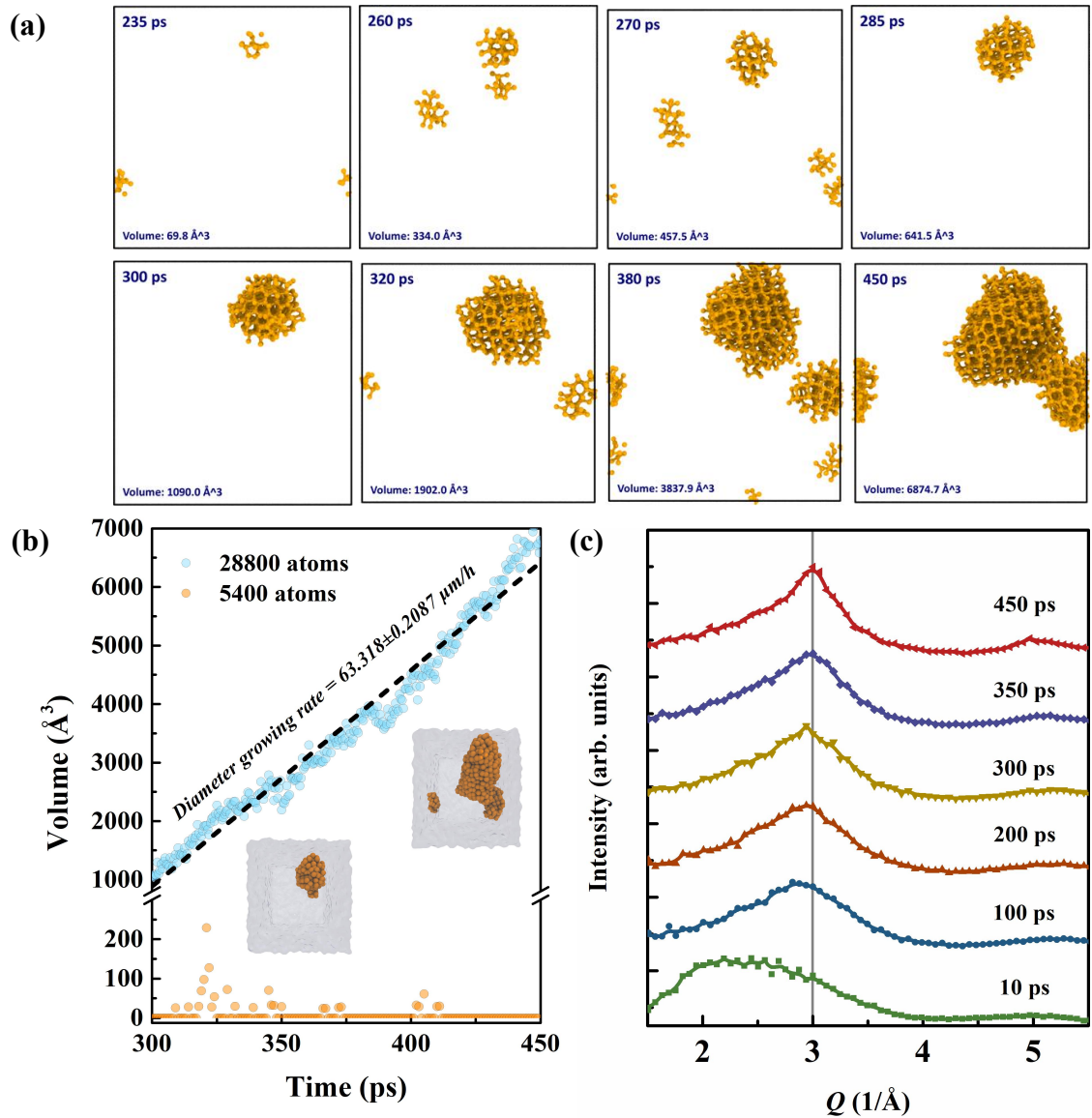

**Fig. 5. Processes of diamonds nucleation under conditions with shock velocity of 15 km/s.** (a) Snapshots of carbon atoms in the diamond structures with computed sizes of 28800 atoms. (b) The temporal evolution of volume in diamonds nucleus. (c) The X-ray diffraction spectra of shock-compressed polyethylene at different simulation times.

## Conclusions

We report the results of large molecular dynamics simulations on the investigation of the dynamic processes in shock-compressed hydrocarbons. A three-step formation of diamonds has been substantiated from decomposition to species separation and nucleation. Bond analysis indicates that splitting decomposition primarily occurs in the CH system and the PE chains break down to give hydrogen and light alkanes including $CH_4$ and $C_2H_6$. In the subsequent hundreds of picoseconds after decomposition, the produced hydrogen-rich molecules escape from the PE chains, leaving the remaining carbon atoms who lost the coordinated hydrogen, bonding with each other. The bonding of carbon atoms and the production of hydrogen molecules promote the separation of the system into two distinct regions with C and H species. In the carbon-rich region, diamond grows spontaneously only if the critical nucleation radius is reached according to the classical nucleation theory. An extremely high diamond growth rate has been found, 2-3 orders of magnitude higher than that achieved using synthesis techniques. A more precise physical diagram of phase separation into hydrogen and diamond in shock-compressed hydrocarbons can be obtained when considering the dynamic processes. These results will improve evolutionary models of heat convection and components in ice giant planets and promote our understanding of critical ablator materials used in inertial confinement fusion experiments.

## Methods

**Simulations of shock compression.** The propagation of shock wave is simulated based on the method of Multi-Scale Shock Technique (MSST)[41]. This methodology combines the Navier-Stokes equations for compressible flow with molecular dynamics simulations. The positions and velocities of all atoms in the system are updated by following the modified Lagrangian for the purpose of restraining the systems at the Hugoniot condition. For the shock wave moving at a speed of $u_s$, the Hugoniot relations can be determined as,

$$u_p = u_s \left(1 - \frac{\rho_0}{\rho}\right) \quad (1)$$

$$p - p_0 = u_s^2 \rho_0 \left(1 - \frac{\rho_0}{\rho}\right) \quad (2)$$

$$e - e_0 = p_0 \left( \frac{1}{\rho_0} - \frac{1}{\rho} \right) + \frac{u_s^2}{2} \left( 1 - \frac{\rho_0}{\rho} \right)^2 \tag{3}$$

Here $u_p$, $\rho$, $p$ and $e$ are the particle velocity, density, pressure, and energy per unit mass respectively. Variables with subscript 0 represent the physical quantities before shock wave. The Lagrangian of atoms are restrained by the conservation equations (1)-(3) to drive the system in a time-independent steady state moving at the shock speed of $u_s$, which can be expressed as,

$$L = T\left(\{\dot{\vec{r}}_i\}\right) - V\left(\{\vec{r}_i\}\right) + \frac{1}{2} Q \dot{\upsilon}^2 + \frac{1}{2} \frac{v_s^2}{\upsilon_0^2} (\upsilon_0 - \upsilon)^2 + p_0 (\upsilon_0 - \upsilon) \tag{4}$$

$$Q \ddot{\upsilon} = \frac{\partial T}{\partial \upsilon} - \frac{\partial V}{\partial \upsilon} - p_0 - \frac{v_s^2}{\upsilon_0^2} (\upsilon_0 - \upsilon) \tag{5}$$

Here $T$, $V$, $Q$, and $\upsilon$ are the kinetic energy per unit mass, the potential energy per unit mass, the mass-like parameter for the simulation cell size, and the specific volume $1/\rho$ respectively. The simulations of shockwave are realised by dynamically adjusting the uniaxial strain of the box until the system reaching in a shock equilibrium state.

**Computational details.** MD simulations are performed by using MSST method implemented under LAMMPS framework[42]. As shown in Fig. S1(a) and (b), the initial configuration of simulations is crystal polyethylene while the shock compression is set along the *z*-axis. The small system contains 5400 atoms with a computational cell size of 2.5nm(*x*)×2.2nm(*y*)×7.6nm(*z*), and the large system contains 28800 atoms with a computational cell size of 4.4nm(*x*)×4.9nm(*y*)×10.2nm(*z*). Periodic boundary conditions are utilized in all directions. All simulations are relaxed at 300 K and 1 bar lasting for 125ps with a timestep of 0.25 fs by using Nose-Hoover thermostat and Parrinello-Rahman barostat. After equilibrium, a series of MSST simulations is performed for 500ps at shock velocities ranging from 2.0~15.0 km/s, generating the shock pressure from 8~125 GPa. We estimate the mass-like parameter $Q$ as 3600, artificial viscosity *mu* as 0.908, and converting factor *tscale* as 0.02 to configure the MSST simulations.

These parameters are set to help equilibrate to the shock Hugoniot and Rayleigh line more rapidly, while they are independent to the results. The integration time step is changed to 0.05 fs for the cases of shock velocity larger than 10 km/s, because the hydrogen atoms vibrate violently at high temperature.

We use a reactive force field (ReaxFF)[31] interatomic potential in MD simulations. The ReaxFF potential can accurately model the covalent and electrostatic interactions for hydrocarbon materials and describe the process of reactive interactions among atoms during shock compression. It bases on a bond-order (BO) formalism in conjunction with charge descriptions to determine the energy contributions. Bond order is calculated directly from interatomic distance by using the empirical formula,

$$BO_{ij} = BO_{ij}^{\sigma} + BO_{ij}^{\pi} + BO_{ij}^{\pi\pi}$$
$$= \exp\left[p_{bo1}\left(\frac{r_{ij}}{r_0}\right)^{p_{bo2}}\right] + \exp\left[p_{bo3}\left(\frac{r_{ij}}{r_0}\right)^{p_{bo4}}\right] + \exp\left[p_{bo5}\left(\frac{r_{ij}}{r_0}\right)^{p_{bo6}}\right]. \quad (6)$$

BO is the bond order between atoms *i* and *j*, and the superscripts denote the respective bond characters. $r_{ij}$ is interatomic distance, $r_0$ is the equilibrium bond length, and the $p_{bo}$ terms are empirical parameters. The parameters of ReaxFF are obtained from Strachan *et al.*[32], which is based on large number of *ab initio* calculations, to characterize the atomic interactions of C/H/O/N materials at extreme conditions. The training set contains bond breaking and compression curves for all possible bonds, angle and torsion bending data for all possible cases, as well as crystal data. A charge equilibration scheme by using QeQ method is applied at each timestep to calculate partial atomic charges and Coulombic interactions.

**Neural network representation of the potential energy surface.** A neural network (NN) based scheme for constructing accurate force fields, which can describe the free energy surface according to the machine learning from *ab initio* data at finite temperature, called deep potential (DP). The DP models used in the simulations are generated with DeePMD-kit packages[43]. To cover a wide range of thermodynamics conditions with little computational consumption as possible, a concurrent learning scheme, Deep Potential Generator (DP-GEN)[44], has been adopted to sample the most compact and adequate data set that guarantees the uniform accuracy of DP in the explored configuration space. We sampled pure hydrogen system (molecular/atomic), pure carbon system (diamond/amorphous), and hydrocarbon system with different chemical composition for a wide range of thermodynamic conditions that covers the regime of single-shock and double shock Hugoniots, quantitatively. The temperature ranges from 300 kelvin to 8000 kelvin, and pressure ranges from ambient condition to 350 GPa.

The embedding network is composed of three layers (25, 50, and 100 nodes) while the fitting network has three hidden layers with 240 nodes in each layer. The radius cutoff is chosen tqo be 5.0 Angstrom. The weight parameters of energy, force and virials in loss function are set to (0.02, 1000, 0.02) at the beginning of training and linearly change toward (1.0, 1.0, 1.0) as the learning rate decays. The density functional theory calculations are all performed with the VASP package[45]. We adopt the strongly constrained and appropriately normed (SCAN) meta-GGA exchange-correlation (XC) functional[46, 47], which can provide a much more accurate description of the H-bond network in hydrocarbon system. We also use van der Waals density functional rVV10 to account for the dispersion interaction of electrons for dense hydrogen[48, 49]. The pseudopotential takes the projector augmented-wave (PAW) formalism[50], and the sampling of Brillouin zone is chosen as 0.5 inverse of angstrom to give accurate virial tensor.

**Structure analysis.** We identify the diamond structure for each atom through OVITO software[51]. The structure identification method can be considered as an extended version of the common neighbour analysis (CNA). Both nearest and second nearest neighbours of an atom can be identified, along with the obtaining of the CNA fingerprints. The central atom was classified as cubic or hexagonal diamond when it is analogously determined as FCC structure or HCP structure respectively. This structure identification method has been described in the work of Maras *et al.*[52]

**Classical nucleation theory.** For a solid nucleus composed of *n* atoms growing spontaneously in a liquid system, the free energy $\Delta G(n)$ must be <0,

$$\Delta G(n) = -\Delta \mu n + \sigma n^{2/3} \tag{7}$$

and

$$\sigma = (36\pi)^{1/3} \rho_s^{-2/3} \gamma, \tag{8}$$

where $\Delta\mu$ is the difference between the chemical potentials of the solid and liquid phases, $\rho_s$ is the number density of the solid cluster, and $\gamma$ is the interfacial free energy. This framework indicates that growth of a crystalline nucleus requires appropriate thermodynamic conditions and certain interfacial energy barrier. When determining the free energy $\Delta G$ as a function of $n$, the critical dimension can be estimated by:

$$N_c = \frac{32\pi\gamma^3}{3\rho_s |\Delta\mu|^3}. \tag{9}$$

Under the shock conditions with shock velocity of 15 km/s, $\gamma$ was estimated to be 1.86 J/m² and $\Delta\mu$ was estimated to be $|\Delta\mu/k_B T| = 0.60$ using data published by Ghiringhelli *et al.*[39]

## Acknowledgements

We are grateful for the insightful discussions of deep potential with Prof Han Wang. This work was supported by the National Key R&D Program of China under Grant No. 2017YFA0403200, the National Natural Science Foundation of China under Grant Nos. 11774429, No. 12047561, and No. 12104507, the NSAF under Grant No. U1830206, the Science and Technology Innovation Program of Hunan Province under Grant No. 2021RC4026 and No. 2020RC2038. All calculations were carried out at the Research Center of Supercomputing Application at NUDT.

## Author contributions

B.C. and J.D. suggested the specific scientific problem and the general idea on methodology, B.C. and J.D. performed the MD simulations and analyzed data, B.C., Q.Z., X. Y. and J. D. interpreted the results, B.C. and J.D. wrote the paper, and Q.Z., X.Y., J.C., S.Z., and D.K. edited the manuscript before submission.

## Competing interests

The authors declare no competing interests.

# Additional information

**Supplementary information** is available for this paper at www.